\def\year{2021}\relax
\documentclass[letterpaper]{article} % DO NOT CHANGE THIS
\usepackage{aaai21}  % DO NOT CHANGE THIS
\usepackage{times}  % DO NOT CHANGE THIS
\usepackage{helvet} % DO NOT CHANGE THIS
\usepackage{courier}  % DO NOT CHANGE THIS
\usepackage[hyphens]{url}  % DO NOT CHANGE THIS
\usepackage{graphicx} % DO NOT CHANGE THIS
\usepackage{graphicx}  % DO NOT CHANGE THIS
\usepackage{natbib}  % DO NOT CHANGE THIS AND DO NOT ADD ANY OPTIONS TO IT
\usepackage{caption} % DO NOT CHANGE THIS AND DO NOT ADD ANY OPTIONS TO IT
\usepackage{booktabs}       % professional-quality tables
\usepackage{makecell}
\usepackage{adjustbox}
\usepackage{algorithm}
\usepackage[noend]{algpseudocode}
\usepackage{amssymb}
\usepackage{mathtools}
\title{Multi-SpectroGAN: High-Diversity and High-Fidelity Spectrogram Generation with Adversarial Style Combination for Speech Synthesis}
\author{% %Authors
    Sang-Hoon Lee\textsuperscript{\rm 1}, Hyun-Wook Yoon\textsuperscript{\rm 2}, Hyeong-Rae Noh\textsuperscript{\rm 1}, Ji-Hoon Kim\textsuperscript{\rm 3}, Seong-Whan Lee\textsuperscript{\rm 1,3}
}
\affiliations{
    \textsuperscript{\rm 1}Department of Brain and Cognitive Engineering, Korea University, Seoul, Korea\\
    \textsuperscript{\rm 2}Department of Computer and Radio Communications Engineering, Korea University, Seoul, Korea\\
    \textsuperscript{\rm 3}Department of Artificial Intelligence, Korea University, Seoul, Korea\\
    \{sh\_lee, hw\_yoon, hr\_noh, jihoon\_kim, sw.lee\}@korea.ac.kr
}

\begin{document}
\maketitle
\begin{abstract}
While generative adversarial networks (GANs) based neural text-to-speech (TTS) systems have shown significant improvement in neural speech synthesis, there is no TTS system to learn to synthesize speech from text sequences with only adversarial feedback. Because adversarial feedback alone is not sufficient to train the generator, current models still require the reconstruction loss compared with the ground-truth and the generated mel-spectrogram directly. In this paper, we present Multi-SpectroGAN (MSG), which can train the multi-speaker model with only the adversarial feedback by conditioning a self-supervised hidden representation of the generator to a conditional discriminator. This leads to better guidance for generator training. Moreover, we also propose  adversarial style combination (ASC) for better generalization in the unseen speaking style and transcript, which can learn latent representations of the combined style embedding from multiple mel-spectrograms. Trained with ASC and feature matching, the MSG synthesizes a high-diversity mel-spectrogram by controlling and mixing the individual speaking styles (e.g., duration, pitch, and energy). The result shows that the MSG synthesizes a high-fidelity mel-spectrogram, which has almost the same naturalness MOS score as the ground-truth mel-spectrogram.
\end{abstract}

\section{Introduction}
Recently, there has been a significant progress in the end-to-end text-to-speech (TTS) model, which can convert a normal text into speech. When synthesizing speech, the recently proposed methods use additional speech audio as an input to reflect the style features from the input audio to the synthesized audio \cite{wang2018style, SkerryRyan2018TowardsEP}. However, there are limitations to transferring and controlling the style without a large amount of high-quality text-audio data (e.g., audiobook dataset). Moreover, because it is difficult to acquire high-quality data, some studies use the knowledge distillation method to improve the performance \cite{ren2019fastspeech}. However, knowledge distillation makes the training complicated, and the generated mel-spectrogram is not complete unlike the ground-truth mel-spectrogram \cite{ren2020fastspeech}.

For better generalization, the current models are trained with adversarial feedback. These generative adversarial networks (GANs) \cite{goodfellow2014generative} based TTS models demonstrate that adversarial feedback is important for learning to synthesize high-quality audio. MelGAN \cite{kumar2019melgan} successfully converts mel-spectrograms to waveforms using a window-based discriminator. The Parallel WaveGAN (PWG) \cite{yamamoto2020parallel} also converts mel-spectrograms to raw waveforms using the adversarial feedback of audio with multi-resolution spectrogram losses. The GAN-TTS \cite{binkowski2019high} also generates raw speech audio with GANs conditioning features that are predicted by separate models. The EATS \cite{donahue2020end} generates the raw waveform from raw phoneme inputs, which is learned end-to-end with various adversarial feedbacks and prediction losses. However, these methods have not yet learned the model without the prediction loss.
\begin{figure*}[t]
\centering
\includegraphics[width=1.0\textwidth]{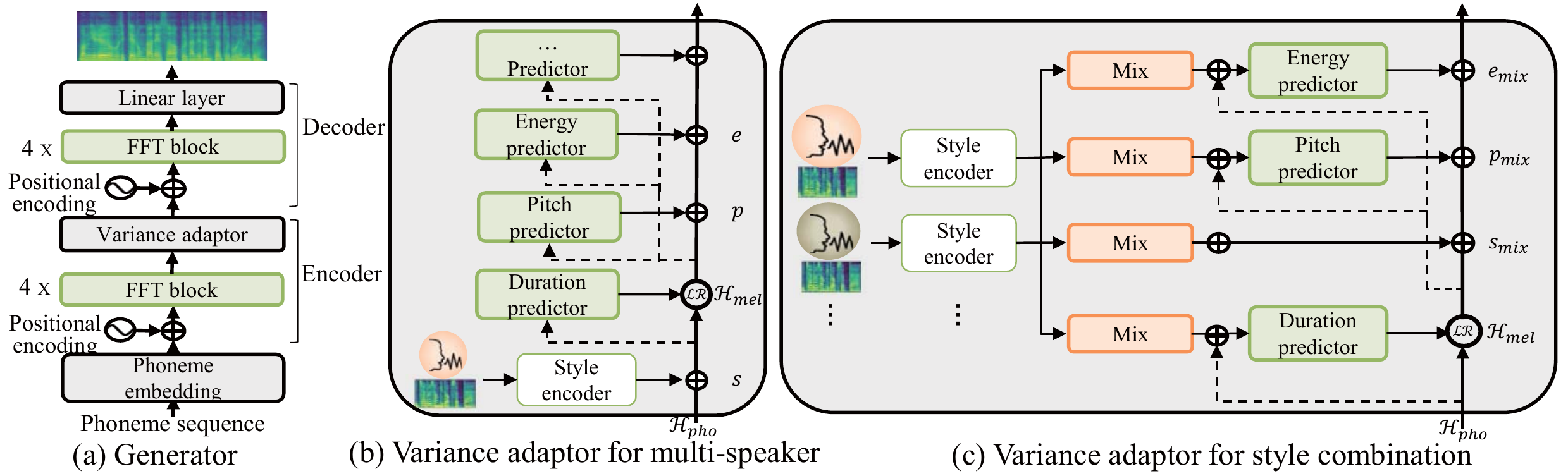} % Reduce the figure size so that it is slightly narrower than the column.
\caption{Generator and the variance adaptor architecture for style combination}
\label{fig1}
\end{figure*}

In this paper, we present the Multi-SpectroGAN (MSG), which can generate high-diversity and high-fidelity mel-spectrograms with adversarial feedback. We introduce an end-to-end learned frame-level condition and conditional discriminator to train the model without prediction loss between ground-truth and generated mel-spectrogram. By making the discriminator learn to distinguish which features are converted to mel-spectrogram with a frame-level condition, the generator is trained with frame-level adversarial feedback to synthesize high-fidelity mel-spectrograms. We also propose the adversarial style combination, which can learn the latent representations of mel-spectrograms synthesized with the mixed speaker embeddings. By training with adversarial feedback from the mixed-style mel-spectrogram, we demonstrate that the MSG synthesizes a more diverse mel-spectrogram by interpolation of multiple styles and synthesizes more natural audio of the unseen speaker. The main contributions of this study are as follows:     
\begin{itemize}
    \item Through an end-to-end learned frame-level condition and conditional discriminator, our model can learn to synthesize mel-spectrogram without prediction loss. 
    \item We propose adversarial style combination, which learns the mixed style of mel-spectrogram with adversarial feedback. 
    \item The MSG achieves a mean opinion score (MOS) of 3.90 with a small amount of multi-speaker data and almost the same MOS with ground-truth mel-spectrogram in single speaker model.
\end{itemize}

\section{Related Works}
\paragraph{Text-to-speech}Autoregressive models such as Tacotron \cite{wang2017tacotron, shen2018natural} were proposed to generate mel-spectrograms through an attention-based recurrent neural network (RNN) \cite{bulthoff2003biologically}. In this model, each frame is autoregressively generated through a sequential generative model conditioned on previously generated frames. However, this method is slow in inference, and it is difficult to model long-term dependencies, leading to word skipping or repetition problems.

To solve these problems, several non-autoregressive models have been proposed for faster generation. FastSpeech \cite{ren2019fastspeech} adapted a feed-forward block from Transformer \cite{vaswani2017attention} with a self-attention mechanism to perform parallel generation. In addition, the model implemented a length regulator to properly match the character-level sequence with the frame-level sequence. FastSpeech2 \cite{ren2020fastspeech} strengthens their model with additional variance information to predict acoustic features more accurately. In FastPitch \cite{lancucki2020fastpitch}, the author cascades fundamental frequency on the phoneme hidden representation \cite{lee1999integrated, yang2007reconstruction}.

With the improved performance of the speech synthesis model, several models have been proposed to control the speaking style of generated speech. One well-known method is the global style token (GST) \cite{wang2018style}, which makes the model learn a prosodic aspect of the variable-length audio signal through several style tokens without any style label. A variational autoencoder (VAE)-based style control model \cite{zhang2019learning} was also proposed while maintaining unsupervised learning in style features.
% Unlike the attempt of latent disentanglement, a transformer-based mode l\cite{dai2019style} was proposed to control speaker style without disentangling latent representation. 
% Instead of internally disentangle style feature in latent space, the model externally injected style feature during the encoder process to conditionally estimate the probability output.

In the Transformer-based TTS model \cite{li2019neural}, training a model with various speakers is challenging because of the difficulty in learning the text-to-speech alignment. \cite{li2020robutrans, chen2020multispeech} identified that the limitation of using location-sensitive attention in the parallel computational model pose a difficulty for the Transformer-based model to learn the alignment between the linguistic and acoustic features. To solve this issue, \cite{chen2020multispeech} used diagonal constraints in encoder-decoder attention to make the model forcefully learn the diagonal area.

\paragraph{Waveform generation}Most speech synthesis models generate intermediate features such as mel-spectrograms to reduce computational time. Therefore, an additional module, named `vocoder', is needed to generate a fully audible signal. %The neural vocoder is categorized into an autoregressive model and a non-autoregressive model.
In an autoregressive model such as Wavenet \cite{oord2016wavenet}, each audio sample is generated sequentially, usually conditioned on previous samples. In general, an RNN-based vocoder, such as bidirectional-RNN or gated recurrent unit (GRU) is used; therefore, the model can predict each sample precisely without long-range constraint dependency. However, owing to the sequential generation process, the overall inference time is slow. Therefore, generating audio samples simultaneously is necessary. 

For parallel generation models, non-autoregressive generation methods such as knowledge distillation \cite{oord2018parallel} and flow-based generative models \cite{prenger2019waveglow,kim2018flowavenet} have been proposed. These models can generate audio samples in parallel, but they suffer from relatively degraded generation quality. Therefore, the issue of improving audio quality has arisen in the parallel generation model. \cite{yoon2020audio}. Recently, the use of GANs \cite{yamamoto2020parallel} to generate high-quality audio in real-time has shown remarkable performance in the field. However, the problem remains when the model is extended to the multi-speaker domain. Therefore, reducing inference time while maintaining audio quality is still a challenging task. Several attempts have been made to fully generate audio waveforms from text input. \cite{binkowski2019high} used various linguistic features including duration and pitch information, to produce high-fidelity audio. \cite{donahue2020end} proposed a novel aligner, which can align between text and mel-frames in parallel.
\paragraph{Mixup}
Mixup was proposed to regularize the neural networks by training the model on convex combination of example-label pairs \cite{zhang2017mixup}. \cite{verma2019manifold} proposed training the model on interpolations of hidden representation. The method for learning combined latent representation of autoencoder was proposed \cite{beckham2019adversarial}. These methods improve the model to generalize for new latent representation which are not seen during training.           

\section{Multi-SpectroGAN}
Our goal is to learn a generator which can synthesize high-diversity and high-fidelity mel-spectrograms by controlling and mixing the speaking style. For high-diversity mel-spectrograms, we introduce an adversarial style combination which can learn latent representations of the combined speaker embedding from multiple mel-spectrograms. To learn the generated mel-spectrogram with randomly mixed styles which doesn’t have a ground truth mel-spectrogram, we propose an end-to-end learned frame-level conditional discriminator. It is also important for better guidance to make the model learn to synthesize speech with only adversarial feedback. We describe the details of the Multi-SpectroGAN architecture and adversarial style combination in the following subsections. 
\subsection{Generator}
We use FastSpeech2 \cite{ren2020fastspeech} as a generator consisting of a phoneme encoder with the variance adaptor denoted as $f(\cdot, \cdot)$, and decoder $g(\cdot)$. We use the phoneme encoder and decoder which consists of 4 feed-forward Transformer (FFT) blocks. Extending to the multi-speaker model, we introduce a style encoder that can produce a fixed-dimensional style vector from a mel-spectrogram like Figure 1.         
\paragraph{Style encoder}
The style encoder has a similar architecture to the prosody encoder of \cite{SkerryRyan2018TowardsEP}. Instead of 2D convolutional network with 3$\times$3 filters and 2$\times$2 stride, our style encoder uses a 6-layer 1D convolutional network with 3$\times$1 filters and 2$\times$2 stride, dropout, ReLU activation, and Layer normalization \cite{ba2016layer}. We also use a gated recurrent unit \cite{cho2014learning} layer and take the final output to compress the length down to a single style vector. Before conditioning the length regulator and variance adaptor, the output is projected as the same dimension of the phoneme encoder output to add style information, followed by a tanh activation function. We denote the style encoder as $E_s(\cdot)$, which produces the style embedding 
\begin{equation}
    \boldsymbol{s} = E_s(\boldsymbol{y}),
 \end{equation}
where $\boldsymbol{s}$ refers to the style embedding extracted from the mel-spectrogram $\boldsymbol{y}$ through the style encoder $E_s$.
%\paragraph{Duration Extractor}
%Because it is hard to learn text-speech alignment in multi-speaker Transformer TTS model, we extract phoneme duration from an encoder-decoder attention alignment of Multi-speaker Tacotron2[] with speaker encoder[]. However, single attention model may focus some phoneme with very large attention, which can make phoneme missing problem where some phonemes have zero length of duration, and it cause length regulator to remove transcript embedding of some phoneme. Our duration extractor forces every phoneme to have at least one length and counts out the duration except for the previous counted phonemes. We describe the details of our duration extractor in supplementary. 
\paragraph{Style-conditional variance adaptor}
With the exception of using style conditional information for learning the multi-speaker model, we use the same variance adaptor of FastSpeech2 \cite{ren2020fastspeech} to add variance information. By adding the style embedding predicted from the mel-spectrogram to the phoneme hidden sequence $\mathcal{H}_{pho}$, the variance adaptor predicts each variance information with the unique style of each speaker. For details, we denote the phoneme-side FFT networks as phoneme encoder $E_p(\cdot)$, which produces the phoneme hidden representation 
\begin{equation}
    \mathcal{H}_{pho}=E_p(\boldsymbol{x}+PE(\cdot)), 
 \end{equation}
where $\boldsymbol{x}$ is the phoneme embedding sequence, and $PE(\cdot)$ is a triangle positional embedding \cite{li2019neural} for giving positional information to the Transformer networks. We extract the target duration sequences $\boldsymbol{\mathcal{D}}$ from Tacotron2 to map the length of the phoneme hidden sequence to the length of the mel-spectrogram
\begin{equation}
    \mathcal{H}_{mel}=\mathcal{LR}(\mathcal{H}_{pho}, \boldsymbol{\mathcal{D}}).
 \end{equation}
The duration predictor predicts the log-scale of the length with the mean-square error (MSE)
\begin{equation}
    \mathcal{L}_{Duration} = \mathbb{E} [\lVert log(\boldsymbol{\mathcal{D}}+1)-\boldsymbol{\mathcal{\hat{D}}} \rVert_2],
 \end{equation}
where % $\hat{D}$ is the log-scale of predicted duration
\begin{equation}
    \boldsymbol{\mathcal{\hat{D}}}=DurationPredictor(\mathcal{H}_{pho}, \boldsymbol{s}).
 \end{equation}
We also use the target pitch sequences $\boldsymbol{\mathcal{P}}$ and target energy sequences $\boldsymbol{\mathcal{E}}$ for each mel-spectrogram frame. We remove the outliers of each information and use the normalized value. Then we add the embedding of quantized $F0$ and energy sequences, $\boldsymbol{p}$ and $\boldsymbol{e}$, which are divided by 256 values.
\begin{equation}
    \boldsymbol{p}=PitchEmbedding(\boldsymbol{\mathcal{P}}), \, \boldsymbol{e}=EnergyEmbedding(\boldsymbol{\mathcal{E}}).
 \end{equation}
The pitch/energy predictor predicts the normalized $F0$/energy value with the MSE between the ground-truth $\boldsymbol{\mathcal{P}}$, $\boldsymbol{\mathcal{E}}$ and the predicted $\boldsymbol{\mathcal{\hat{P}}}$, $\boldsymbol{\mathcal{\hat{E}}}$
\begin{equation}
\begin{split}
    \mathcal{L}_{Pitch} =  \mathbb{E}[\lVert \boldsymbol{\mathcal{P}}-\boldsymbol{\mathcal{\hat{P}}} \rVert_2],
\\
    \mathcal{L}_{Energy} =  \mathbb{E}[\lVert \boldsymbol{\mathcal{E}}-\boldsymbol{\mathcal{\hat{E}}} \rVert_2],
\end{split}
 \end{equation}
where 
\begin{equation}
\begin{split}
    \boldsymbol{\mathcal{\hat{P}}}=PitchPredictor(\mathcal{H}_{mel},\boldsymbol{s}), 
\\
    \boldsymbol{\mathcal{\hat{E}}}=EnergyPredictor(\mathcal{H}_{mel},\boldsymbol{s}).
\end{split}
 \end{equation}
The encoder $f(\cdot, \cdot)$ consisting of a phoneme encoder and style-conditional variance adaptor is trained with the variance prediction loss     
\begin{equation}
   \min_{f} \mathcal{L}_{var} = \mathcal{L}_{Duration}+\mathcal{L}_{Pitch}+\mathcal{L}_{Energy}.
 \end{equation}
During training, we use not only the ground-truth value of each information, such as \cite{ren2020fastspeech}, but also the predicted value of each information with adversarial style combination to learn the variety of generated mel-spectrograms without the ground-truth.    
The sum of each informational hidden sequence $\mathcal{H}_{total}$ is passed to the decoder as a generator $g(\cdot)$ to generate a mel-spectrogram as
\begin{equation}
    \mathcal{H}_{total} = \mathcal{H}_{mel}+\boldsymbol{s}+\boldsymbol{p}+\boldsymbol{e}+PE(\cdot),
 \end{equation}
\begin{equation}
    \boldsymbol{\hat{y}}=g(\mathcal{H}_{total}),
 \end{equation}
where $\boldsymbol{\hat{y}}$ is the predicted mel-spectrogram. Our baseline models use the reconstruction loss with mean-absolute error (MAE) as
\begin{equation}
    \mathcal{L}_{rec}=\mathbb{E} [\lVert \boldsymbol{y}-\boldsymbol{\hat{y}}\rVert_1], 
 \end{equation}
where $\boldsymbol{y}$ is the ground-truth mel-spectrogram.

\begin{figure*}[t]
\centering
\includegraphics[width=1.0\textwidth]{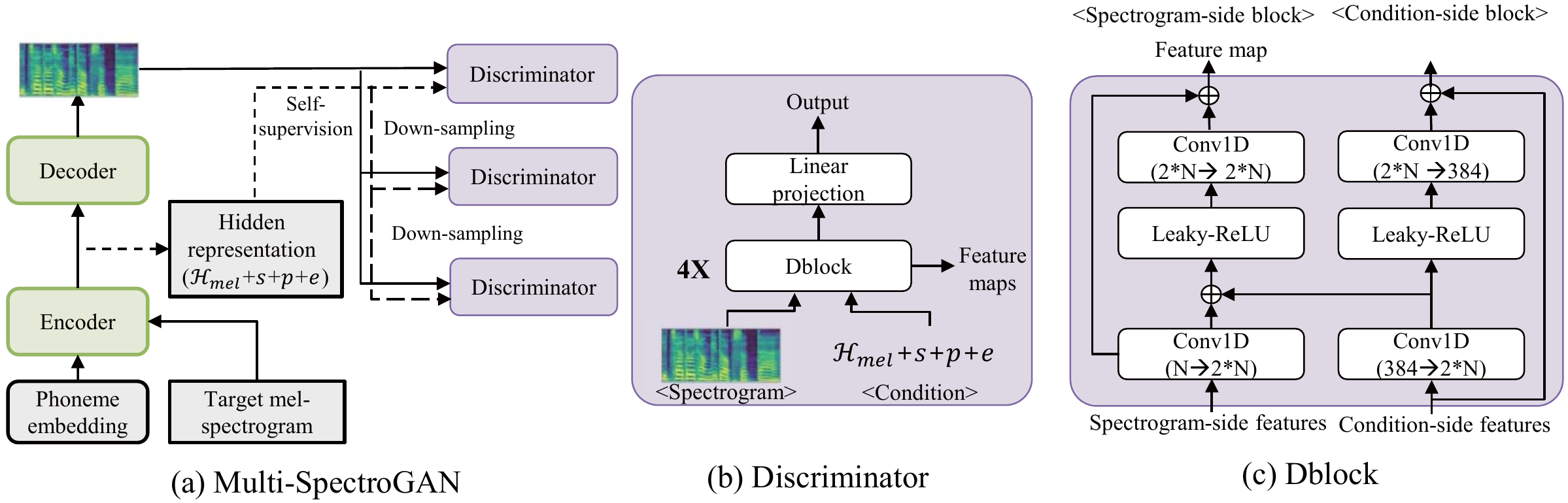} % Reduce the figure size so that it is slightly narrower than the column.
\caption{Frame-level conditional discriminator. Each discriminator has 4 Dblocks consisting of spectrogram-side block and condition-side block. Each side has two non-strided 1D convolutional networks with kernel size of 3. Conditional hidden states are added to spectrogram-side hidden states by the same filter size after first convolutional layer.}
\label{fig2}
\end{figure*}

\subsection{Discriminator}
Unlike the previous GAN-based TTS model, our model can be learned to synthesize the mel-spectrogram from a text sequence without calculating the loss compared with the ground-truth spectrogram directly. To train the model without $\mathcal{L}_{rec}$, we design a frame-level conditional discriminator using the end-to-end learned frame-level condition.
\paragraph{End-to-end learned frame level condition}
To learn to distinguish between the frame-level real and generated mel-spectrogram, the discriminator uses the encoder outputs as a frame-level condition that is learned in a generator during training. Note that $\boldsymbol{c}$ is the sum of linguistic, style, pitch, and energy information, which is end-to-end learned in a generator during training and is expressed as:    
\begin{equation}
    \boldsymbol{c} =  \underbrace{\mathcal{H}_{mel}}_\text{linguistic}+\underbrace{\boldsymbol{s}}_\text{style}+\underbrace{\boldsymbol{p}}_\text{pitch}+\underbrace{\boldsymbol{e}}_\text{energy}.
 \end{equation}
\paragraph{Frame-level conditional discriminator}
As shown in Figure 2, we adopt a multi-scale discriminator that has identical network structure like MelGAN \cite{kumar2019melgan}. While MelGAN motivates the multiple discriminators at different scales to learn features for the different frequency ranges of the audio, we choose multiple discriminators to learn features for different ranges of linguistic, pitch, and energy information. 
Each discriminator consists of 4 Dblocks that have a mel-spectrogram side block and a condition side block. Each block uses a 2-layer non-strided 1D convolutional network with the Leaky-ReLU activation function to extract the adjacent frame information. We add the hidden representation of the condition side block to the mel-spectrogram side hidden representation. Similar to \cite{vaswani2017attention}, residual connections and layer normalization is used at each block output for optimization.

We use the least-squares GAN (LSGAN) \cite{mao2017least} formulation to train the Multi-SpectroGAN. The discriminators $D_k$ learn to distinguish between real spectrogram $\boldsymbol{y}$ and reconstructed one from $\boldsymbol{x}$, $\boldsymbol{y}$. We minimize the GAN loss from the mel-spectrogram. The encoder $f(\cdot, \cdot)$ and decoder $g(\cdot)$ as a generator, and discriminator $D$ are trained by the following losses:  
\begin{equation}
\begin{split}
    \min_{D_{k}}\mathbb{E}[{\rVert D_{k}(\boldsymbol{y}, \boldsymbol{c})-1 \rVert_2}+\rVert D_{k}(\boldsymbol{\hat{y}}, \boldsymbol{c})\rVert_2], \forall k=1,2,3
\end{split}
 \end{equation}
\begin{equation}
    \mathcal{L}_{adv}= \mathbb{E} \left[\sum_{k=1}^{3}\rVert D_{k}(\boldsymbol{\hat{y}}, \boldsymbol{c})-1\rVert_2\right].
 \end{equation}
\paragraph{Feature matching}
To improve the representations learned by the discriminator, we use a feature matching objective like \cite{kumar2019melgan}. Unlike the MelGAN, which minimizes the MAE between the discriminator feature maps of real and generated audio, we minimize the MAE between the feature maps of each spectrogram-side block:
% Note that the $\mathcal{L}_{rec}$ which are used in raw mel-spectrogram space directly make it hard to learn the discriminator.
\begin{equation}
    \mathcal{L}_{fm} = \mathbb{E}\left[\sum_{i=1}^{4}\frac{1}{N_i} \lVert D_{k}^{(i)}(\boldsymbol{y}, \boldsymbol{c} )-D_{k}^{(i)}(\boldsymbol{\hat{y}}, \boldsymbol{c}) \rVert_1\right], 
 \end{equation}
where $D_{k}^{(i)}$ refers to the $i^{th}$ spectrogram-side block output of the $k^{th}$ discriminator,  and $N_{i}$ is the number of units in each block output.
The generator trains with the following objective:
\begin{equation}
  \min_{f, g} \mathcal{L}_{msg} = \mathcal{L}_{adv}+\lambda\mathcal{L}_{fm}+\mu\mathcal{L}_{var}.
 \end{equation}

\subsection{Adversarial Style Combination}
By introducing the adversarial loss, we would like to synthesize a more realistic audio signal with high-fidelity generated mel-spectrogram. In addition, our goal is to generate a more diverse audio signal with an even unseen style. To do this, we propose the adversarial style combination (ASC), which can make the mel-spectrogram more realistic with the mixed style of multiple source speakers. Similar to \cite{beckham2019adversarial} interpolating the hidden state of the autoencoder for adversarial mixup resynthesis, we use two types of mixing, binary selection between style embeddings, and manifold mixup \cite{verma2019manifold} by the linear combination of style embeddings from the different speakers:    
\begin{equation}
 s_{mix} = \alpha \boldsymbol{s}_i+(1-\alpha)\boldsymbol{s}_j,
 \end{equation}
where $\alpha\in\{0,1\}$ is sampled from a Bernoulli distribution in binary selection and $\alpha\in[0,1]$ is sampled from the Uniform(0,1) distribution in manifold mixup. The variance adaptor predicts each information with a mixed style embedding. Unlike pitch and energy, we use the ground-truth $\mathcal{D}$ randomly selected from multiple source speakers because the duration predictor may predict the wrong duration at the early training step. Each variance information is predicted by different ratios of mixed style embedding. We call it ``style combination", in which the final mixed hidden representation is the combination of each variance information from different mixed styles:
% \begin{equation}
%     \hat{P}_{mix}=PitchPredictor(\mathcal{H}_{mel},s_{mix}), 
%  \end{equation}
% \begin{equation}
%      \hat{E}_{mix}=EnergyPredictor(\mathcal{H}_{mel},s_{mix}),
%  \end{equation}
% \begin{equation}
%     p_{mix}=PitchEmbedding(\hat{P}_{mix})
%  \end{equation}
% \begin{equation}
%     e_{mix}=EnergyEmbedding(\hat{E}_{mix})
%  \end{equation}
\begin{equation}
    \mathcal{H}_{mix} = \underbrace{\mathcal{H}_{mel}+\boldsymbol{s}_{mix}+\boldsymbol{p}_{mix}+\boldsymbol{e}_{mix}}_\text{$\boldsymbol{c}_{mix}$}+PE(\cdot),
 \end{equation}
\begin{equation}
    \boldsymbol{\hat{y}}_{mix}=g(\mathcal{H}_{mix}),
 \end{equation}
where $\boldsymbol{p}_{mix}$ and $\boldsymbol{e}_{mix}$ are the pitch and energy embedding of the predicted value from mixed styles, respectively, and $\boldsymbol{c}_{mix}$ is fed to discriminator as the frame-level condition for mel-spectrogram $\boldsymbol{\hat{y}}_{mix}$ generated by style combination.   
The discriminator is trained using the following objective:  
\begin{equation}
\begin{split}
    \min_{D_{k}}\mathbb{E} & [{\rVert D_{k}(\boldsymbol{y}, \boldsymbol{c})-1\rVert_2}+\rVert D_{k}(\boldsymbol{\hat{y}}, \boldsymbol{c})\rVert_2\\
    &+\rVert D_{k}(\boldsymbol{\hat{y}}_{mix}, \boldsymbol{c}_{mix})\rVert_2],  \forall k=1,2,3.
    %\min_{D_{k}}\mathbb{E} [{\rVert D_{k}(\boldsymbol{y}, \boldsymbol{c})-1\rVert_2}+\rVert D_{k}(\boldsymbol{\hat{y}}, \boldsymbol{c})\rVert_2
    %+\rVert D_{k}(\boldsymbol{\hat{y}}_{mix}, \boldsymbol{c}_{mix})\rVert_2], \\   \forall k=1,2,3.
 %   \min_{D_{k}}\mathbb{E}[{\rVert D_{k}(\boldsymbol{y}, \boldsymbol{c})-1\rVert_2}+\rVert D_{k}(\boldsymbol{\hat{y}}, \boldsymbol{c})\rVert_2\\+\rVert D_{k}(\boldsymbol{\hat{y}}_{mix}, \boldsymbol{c}_{mix})\rVert_2], \forall k=1,2,3.
\end{split}
\end{equation}
The generator is trained by the following loss: 
\begin{equation}
  \min_{f, g} \mathcal{L}_{asc} = \mathcal{L}_{adv}+\lambda\mathcal{L}_{fm}+\mu\mathcal{L}_{var}+\nu\mathcal{L}_{mix}, 
 \end{equation}
where %$\mathcal{L}_{mix}$ is minimized to synthesize realistic mel-spectrogram with mixed style by the following objective:  
\begin{equation}
    \mathcal{L}_{mix}= \mathbb{E} \left[\sum_{k=1}^{3}\rVert D_{k}(\boldsymbol{\hat{y}}_{mix}, \boldsymbol{c}_{mix})-1\rVert_2\right].
 \end{equation}
\begin{table}
\centering
\begin{tabular}{lcc}
\Xhline{3\arrayrulewidth}
Model                                 & MOS & 95\% CI \\ 
\hline
GT                             & 4.20& $\pm$ 0.03        \\  
GT (Mel + PWG)               & 3.94& $\pm$ 0.03        \\ 
\hline
Transformer TTS (Mel + PWG)        & 3.83& $\pm$ 0.03        \\
\hline
FastSpeech (Mel + PWG)        & 3.52& $\pm$ 0.04       \\ 
FastSpeech2 (Mel + PWG)        & 3.85& $\pm$ 0.03        \\ 
\hline
MSG (Mel + PWG)  & 3.91& $\pm$ 0.03        \\ 
\Xhline{3\arrayrulewidth}
\end{tabular}
\caption{MOS with 95\% CI for a single speaker model}
\end{table}
\section{Experiments and Results}
We evaluated in the single-speaker and multi-speaker dataset. Ablation studies are performed for downsampling size, loss function, and conditional information. We also evaluated the style-combined speech by control and interpolation of multiple styles. 
We used a Nvidia Titan V to train the single-speaker model with the LJ-speech dataset and the multi-speaker model with the VCTK dataset. Each dataset is split into train, validation, and test. Mel-spectrogram is transformed following the work of \cite{shen2018natural} with a window size of 1024, hop size of 256, 1024 points of Fourier transform, and 22,050 Hz sampling rate. We use the ADAM \cite{kingma2014adam} optimizer with $\beta_1$ = 0.9, $\beta_2$ = 0.98, and $\epsilon$ = 10$^{-9}$, and apply the same learning rate schedule as that of \cite{vaswani2017attention} with an initial learning rate of  ${10}^{-4}$ for $f$, $g$, and $D$. The $\lambda$, $\mu$, and $\nu$ are set to 10, 1 and 1. The phoneme sequences were converted using the method of \cite{g2pE2019}. To convert the mel-spectrogram to audio, we use the pretrained PWG vocoder \cite{yamamoto2020parallel} consisting of 30-layers of dilated residual convolution blocks.    
\begin{table}
\centering
\begin{tabular}{lccc}
\Xhline{3\arrayrulewidth}
Model &   $\tau$  & CMOS & Convergence\\ 
\hline
MSG &2   & 0    & 350k    \\  
MSG &3  & +0.07 & 650k    \\ 
MSG &4  & +0.06 & 1,000k   \\ 
\Xhline{3\arrayrulewidth}
\end{tabular}
\caption{CMOS comparison for the down-sampling size}
\end{table}
\begin{table}
\centering
\begin{tabular}{llc}
\Xhline{3\arrayrulewidth}
Model&Loss function& MOS \\ \hline
FastSpeech2&$\mathcal{L}_{var}$+$\mathcal{L}_{rec}$& 3.85 $\pm$ 0.03         \\  
MSG (w/o $\boldsymbol{c}$)&$\mathcal{L}_{var}$+$\mathcal{L}_{adv}$& -       \\  
MSG (w/ $\boldsymbol{c})$&$\mathcal{L}_{var}$+$\mathcal{L}_{adv}$& 3.14 $\pm$ 0.06     \\ 

MSG (w/ $\boldsymbol{c})$&$\mathcal{L}_{var}$+$\mathcal{L}_{adv}$+$\mathcal{L}_{rec}$& 3.85 $\pm$ 0.03       \\ 
MSG (w/ $\boldsymbol{c})$&$\mathcal{L}_{var}$+$\mathcal{L}_{adv}$+$\mathcal{L}_{rec}$+$\mathcal{L}_{fm}$   & 3.89 $\pm$ 0.03     \\ \hline
MSG (w/ $\boldsymbol{c})$&$\mathcal{L}_{var}$+$\mathcal{L}_{adv}$+$\mathcal{L}_{fm}$ & 3.91 $\pm$ 0.03      \\
\Xhline{3\arrayrulewidth}
\end{tabular}
\caption{Ablation study for the loss function}
\end{table}
\subsection{Single-speaker speech synthesis}
\paragraph{Naturalness MOS}
To evaluate the quality of the synthesized mel-spectrogram, we conducted a subjective MOS test. We randomly selected 100 sentences from the test dataset. The audio generated from each model was sent to Amazon's Mechanical Turk (MTurk). Samples were evaluated by 20 raters on a scale from 1 to 5 with 0.5 point increments. We compared the MSG model with the ground-truth audio (GT), the converted audio from the mel-spectrogram of the GT, and other TTS models using PWG. As shown in Figure 1, the MOS results show that the MSG has an almost similar score to the ground-truth mel-spectrogram, which demonstrates our discriminator and the frame-level conditional information improves voice quality even though the same generator architecture \cite{ren2020fastspeech} is used.  
\paragraph{Down-sampling size}
We use average pooling with different kernel sizes to compare downsampling size $\tau$. The model with a downsampling size of 3 has the highest score. The smaller size of downsampling makes the model converge early step with a -0.07 CMOS score. The larger size of the downsampling causes the model to converge slowly but shows a similar MOS. Therefore, we adopted a downsampling size of 3 for our MSG model.
\begin{table*}[ht]
  \centering
  \begin{tabular}{lllcccc}
    \Xhline{3\arrayrulewidth}
    Model    & Mix & ratio &  MOS     & MCD$_{13}$& $F_{0}$ RMSE  & Top-1 acc. \\
    \hline
    GT  & -&-& 4.11$\pm$0.03 &  - & - & 93\%    \\
    GT (Mel + PWG)      &-&-& 4.00$\pm$0.03  & 4.46 & 43.59 & 84\%  \\
    \hline
    Tacotron2 (Mel + PWG)     &-&- & 3.81$\pm$0.04\ & 5.88  & 44.51 & 75\%\\
    
    GST (Mel + PWG)  &-&-& 3.89$\pm$0.04 & 5.59 & 45.10 & \textbf{80\%} \\
    \hline
    FastSpeech2 (Mel + PWG)    &-&-& 3.81$\pm$0.04  & 5.78  & 46.90 & 67\%\\

    MSG (Mel + PWG)     &-&-& 3.89$\pm$0.04  & 5.59 & 45.71 & 72\% \\

     \hline
    MSG+ASC (Mel + PWG)      &Bern& $\{r, r, r,... \}$ & 3.85$\pm$0.04   & \textbf{5.54}  & 45.36 & 70\%\\
    MSG+ASC (Mel + PWG) & Mixup & $\{r, r, r,... \}$ & 3.89$\pm$0.04  & 5.60 & 45.31&69\%\\
    MSG+ASC (Mel + PWG)      &Bern& $\{r_s, r_p, r_e,... \}$& 3.87$\pm$0.04 & 5.57  & 47.06 & \textbf{79\%}\\
    MSG+ASC (Mel + PWG)    &Mixup& $\{r_s, r_p, r_e,... \}$& \textbf{3.90$\pm$0.04} & 5.57  & \textbf{43.97} & 73\%\\
    \Xhline{3\arrayrulewidth}
  \end{tabular}
  \caption{Results of subjective and objective tests for seen speaker. Bern refers that the ratio is sampled from a Bernoulli distribution. Mixup refers that the ratio is sampled from the uniform (0,1) distribution. We compare the models with same ratios \{$r$, $r$, $r$,...\} and different ratios for mixing the style and each variance \{$r_s$, $r_p$, $r_e$,...\} where $r_s$, $r_p$, and $r_e$ are the ratios for mixing the style, pitch, and energy embeddings respectively.}
\end{table*}
\begin{table*}[ht]
  \centering
  \begin{tabular}{lllcccc}
    \Xhline{3\arrayrulewidth}
    Model    & Mix & ratio &  MOS     & MCD$_{13}$ & $F_{0}$ RMSE& Top-1 acc. \\
    \hline
    GT  & -&-& 4.00$\pm$0.03& - & - & 95\%      \\
    GT (Mel + PWG)      &-&-& 3.96$\pm$0.03  & 4.26 &   49.56 &88\% \\
    \hline
    Tacotron2 (Mel + PWG)     &-&- & 3.76$\pm$0.04  & 6.33  & 46.26 &17\% \\

    GST (Mel + PWG)      &-&-& \textbf{3.83$\pm$0.04}  & 6.15  & 41.71 &5\%\\
    \hline
    FastSpeech2 (Mel + PWG)    &-&-& 3.67$\pm$0.04  & 6.18  & 48.31 &20\%  \\
    MSG (Mel + PWG)     &-&-& 3.80$\pm$0.04  & 6.10  & 48.02 &23\%\\
     \hline
    MSG+ASC (Mel + PWG)     &Bern& $\{r, r, r,... \}$ & 3.80$\pm$0.04  & 6.11  & \textbf{47.04} &\textbf{30\%} \\
    MSG+ASC (Mel + PWG) & Mixup & $\{r, r, r,... \}$ & \textbf{3.82$\pm$0.04}  & \textbf{6.07}  & 47.69 &27\% \\
    MSG+ASC (Mel + PWG)      &Bern& $\{r_s, r_p, r_e,... \}$& 3.75$\pm$0.04  & 6.14  & 48.10 &28\% \\
    MSG+ASC (Mel + PWG)    &Mixup& $\{r_s, r_p, r_e,... \}$ & 3.81$\pm$0.04  & 6.08 & 47.22 &\textbf{30}\%\\
    \Xhline{3\arrayrulewidth}
  \end{tabular}
  \caption{Results of subjective and objective tests for unseen speaker.}
\end{table*}
\paragraph{Loss function}
We conducted the ablation study for the loss functions and the conditional discriminator. When the conditional information of the discriminator is replaced with $z$ noise and trained with the loss function of $\mathcal{L}_{var}$ and $\mathcal{L}_{adv}$, this model does not train at all. On the other hand, the model using conditional information in the discriminator can learn to synthesize the mel-spectrogram without $\mathcal{L}_{rec}$ or $\mathcal{L}_{fm}$ which must be calculated between the ground-truth and generated mel-spectrogram. This demonstrates that the frame-level conditional discriminators using the end-to-end learned frame-level condition make it possible to train the model even if the generated mel-spectrogram does not have ground-truth audio. However, we also use the additional loss function $\mathcal{L}_{rec}$ or $\mathcal{L}_{fm}$ to improve the audio quality. Although most TTS models train with $\mathcal{L}_{rec}$, it is too strong supervision to train with adversarial loss; therefore, adversarial loss has a slight influence on the model. Unlike $\mathcal{L}_{rec}$, the $\mathcal{L}_{fm}$ is affected by the discriminator, and it shows the highest MOS score when the model was trained with $\mathcal{L}_{fm}$.     
\subsection{Multi-speaker speech synthesis}
We trained each model using 30 speakers in the VCTK dataset. We evaluated each model with ``seen speaker" and ``unseen speaker" of reference audio for style. The ``seen speaker" of reference audio indicates the audio of the speaker seen during training. The ``unseen speaker" of reference audio indicates the audio of the speaker unseen during training, which is evaluated for the zero-shot style transfer. Audio samples of the generated speech are provided.\footnote{\url{https://anonymsg.github.io/MSG/Demo/index.html}}
\paragraph{Naturalness MOS}
For the subjective MOS test of each multi-speaker model, we randomly selected 40 speakers (20 seen and 20 unseen speakers) and 5 sentences from a test dataset of each speaker. The samples were evaluated by 20 raters on a scale of 1-5 with 0.5 point increments through Amazon MTurk. We compared our models with GT, the converted audio from the mel-spectrogram of the GT, and other TTS models (Tacotron2, GST, Tansformer-based TTS, and FastSpeech2). For multi-speaker Tacotron2, we add the style encoder and concatenate with the transcript embedding. In a Transformer-based TTS model, it is not possible to synthesize any audio because of the wrong alignment. For multi-speaker FastSpeech2, we train the model with the same style encoder and add the style embedding to transcript embedding. Even though using the same generator structure with FastSpeech2, the results show our method improves the audio quality of 0.08 for seen speaker and 0.13 for unseen speaker. When trained with ASC, the models have better performance on both the seen and unseen speakers.
\begin{figure*}[t]
\centering
\includegraphics[width=1\textwidth]{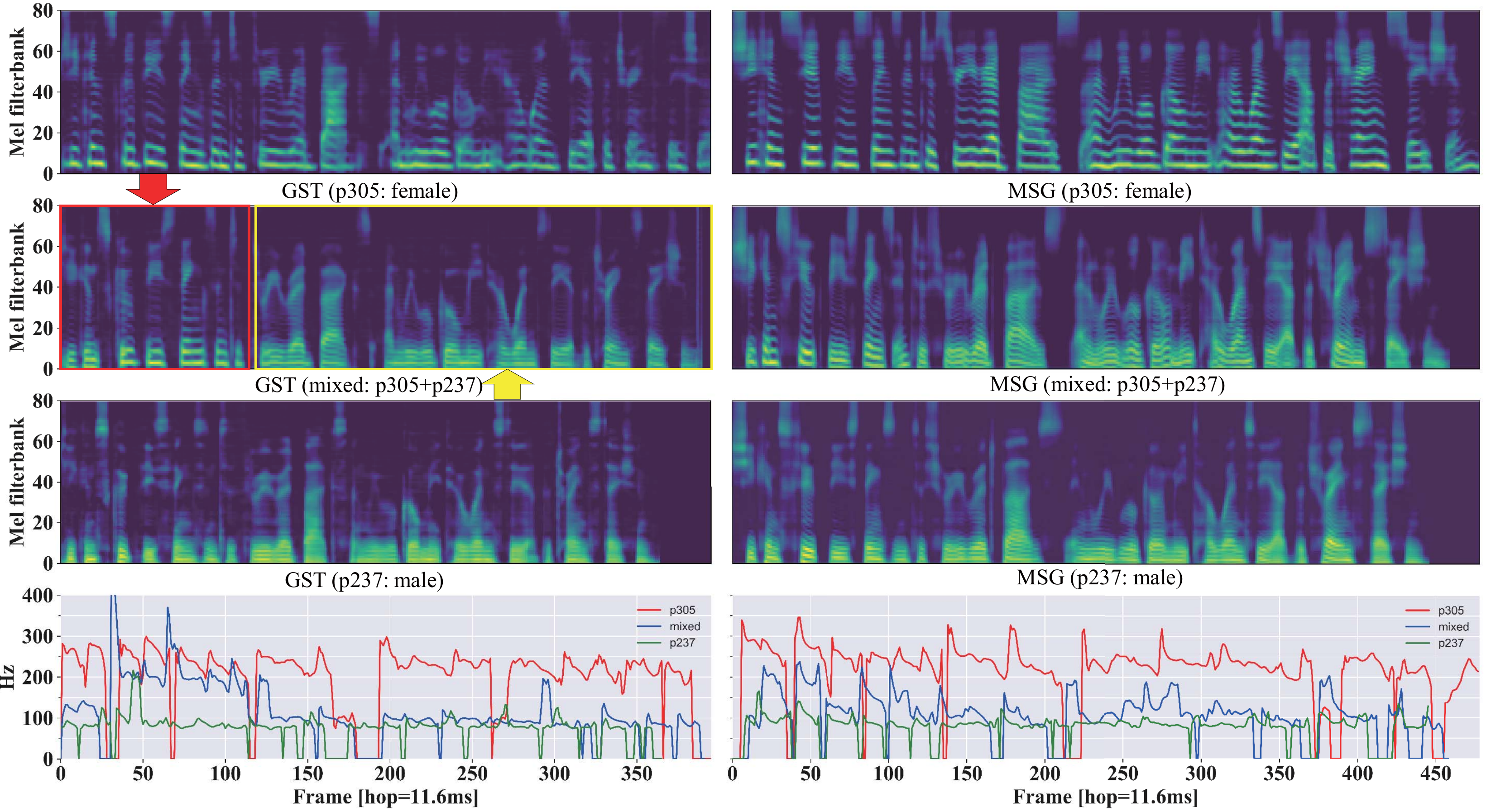} % Reduce the figure size so that it is slightly narrower than the column.
\caption{Mel-spectrogram and $F_0$ contour of the GST (Left) and MSG (Right).}
\label{fig3}
\end{figure*}
\paragraph{Objective evaluation}
We conducted an objective evaluation using mel-cepstral distortion (MCD)  \cite{kubichek1993mel}, $F_0$ root mean squared error (RMSE), and speaker classification \cite{wan2018generalized}. To evaluate each metric, each model synthesized 100 utterances for both the seen and unseen speaker. For comparison of $F_0$ RMSE, we used target duration for FastSpeech2 and our models, and teacher-forcing synthesis with target mel-spectrogram for Tacotron2 and GST. Even though the GST shows the highest MOS score in the unseen speaker, the top-1 speaker classification accuracy is 5\%, where the GST only synthesizes the learned voice during training. When the model is trained with ASC, the results verify that learning the combined latent representation in training makes the model synthesize a more diversed mel-spectrogram even for unseen speakers.
\begin{table}
\centering
\begin{tabular}{lcc}
\Xhline{3\arrayrulewidth}
Condition & Loss function & MOS \\ 
\hline
MSG ($\boldsymbol{c}$) &$\mathcal{L}_{var}$ + $\mathcal{L}_{adv}$       & 3.57$\pm$ 0.07     \\
~~$-\mathcal{H}_{mel}$ &$\mathcal{L}_{var}$ + $\mathcal{L}_{adv}$& [does not train]    \\ 
~~$-\boldsymbol{s}-\boldsymbol{p}$ &$\mathcal{L}_{var}$ + $\mathcal{L}_{adv}$  & [does not train]      \\ 
~~$-\boldsymbol{p}$      &$\mathcal{L}_{var}$ + $\mathcal{L}_{adv}$  & 3.52$\pm$ 0.07       \\ 
~~$-\boldsymbol{e}$   &$\mathcal{L}_{var}$ + $\mathcal{L}_{adv}$    & 3.54$\pm$ 0.07 \\ 
\Xhline{3\arrayrulewidth}
\end{tabular}
\caption{Ablation study for condition of discriminator}
\end{table}

\paragraph{Ablation study}
We conducted an ablation study for the conditions in the discriminator. To evaluate the effectiveness of each conditional information, we trained the model without $\mathcal{L}_{fm}$. The model without $\mathcal{H}_{mel}$ does not train at all, which demonstrates that linguistic information is essential to learn to synthesize the frame-level mel-spectrogram. Unlike a single-speaker model that can learn to synthesize without style $\boldsymbol{s}$ or pitch $\boldsymbol{p}$ information, the multi-speaker model without $\boldsymbol{s}$ and $\boldsymbol{p}$ does not train at all. The model without $\boldsymbol{p}$ and $\boldsymbol{e}$ shows that each information has an effect on naturalness.       
\subsection{Style Combination}
For the robustness of style transfer and control, we synthesize the mel-spectrogram with mixed style embedding which are interpolated style embedding of two speakers (1 male and 1 female). Figure 3 shows the mel-spectrograms and $F0$ contour (women, mixed and men style embedding) of GST (Left) and MSG (Right) model for the same sentence. The attention-based autoregressive models have some problems. Even when using an unseen and mixed style, the models synthesize a mel-spectrogram with a seen style during training. In addition, the change in the voice occurs at the same utterance as in Figure 3. Even in most cases, word skipping and repetition occur because the models fail to align.

Unlike attention-based autoregressive models, the MSG model trained with adversarial style combination synthesizes the mel-spectrogram robustly even with mixed-style embedding. The results demonstrate that the synthesis with the interpolated style embedding can generate a new style of mel-spectrogram by a combination of two styles. We also synthesized a particular style of a mel-spectrogram in combination with the desired proportions of each variance information (e.g., duration, pitch, and energy).

\section{Conclusion and Future Work}
We presented a Multi-SpectroGAN, which can generate high-diversity and high-fidelity mel-spectrograms with adversarial style combination. We demonstrated that it is possible to train the model with only adversarial feedback by conditioning a self-supervised latent representation of the generator to the discriminator. Our results also showed the effectiveness of mixing hidden states in the audio domain, which can learn the mel-spectrogram generated from a combination of mixed latent representations. By exploring various style combination for mixup, we show that learning the mel-spectrogram of mixed style made the model generalize better even in the case of unseen transcript and unseen speaker. For future work, we will train the Multi-SpectroGAN with few-shot learning and cross-lingual style transfer frameworks.
% \end{linenumbers}

\section{Acknowledgments}
This work was supported by Institute for Information \& communications Technology Planning \& Evaluation (IITP) grant funded by the Korea government (MSIT) (No. 2019-0-01371, Development of brain-inspired AI with human-like intelligence \& No. 2019-0-00079,  Artificial Intelligence Graduate School Program, Korea University), Netmarble AI Center, and the Seoul R\&BD Program (CY190019).
\bibliography{aaai}

\end{document}